\documentclass[10pt]{article}

\begin{document}
\title{ Solutions of some  non-Hermitian Schr\"{o}dinger Equations}
 \author { B L Burrows  \\ \\ \small Emeritus Professor  \\ \small{ Staffordshire University, College Road, Stoke-on-Trent,ST4 2DE UK} \\ \small{e-mail:brian.burrows2@btopenworld.com} }

    \date{}

\maketitle
\section {Abstract}
Two types of non-Hermitian systems are considered. One of them is both non-Hermitian and non-Linear and an iterative process is used to obtain excited state solutions; the ground state may be solved exactly. The model has been used in many physical systems and the method of calculation uses a simple Hilbert space with a generalised inner product. The second type has a complex term in the Hamiltonian and is a well studied  problem in the infinite interval. Here a finite interval is considered and a complete set of eigenfunctions for this interval is used.The relationship between the finite interval states and the infinite interval states is discussed. 
\section{Introduction}
There has been much recent interest in \emph{non-Hermitian operators} and a selection of this work can be found in references  [1-12 ]. The definition of any operator necessarily includes the domain and space on which it is defined and there are many advantages in  the calculation  of eigenstates of operators that are Hermitian in the given space. In particular the eigenvalues are real and the eigenvectors may be chosen to be orthonormal. If we consider a Hermitian operator  on a one-dimensional  function space of square integrable functions then the eigenvectors have a nodal structure where the nth  eigenvector  has n nodes and between the nodes of the (n+1)st  there is a node of the nth state. For non-Hermitian some or all of these advantages may be absent.In this paper we examine example  eigenvalue problems for two classes of operators and illustrate ways of overcoming the disadvantages.One of these classes is both non-Hermitian and also nonlinear and leads to additional  problems in the necessary numerical approximation of the required integrals and in the convergence of the iteration process. The second class of potentials has imaginary terms in the corresponding potential. This system has been considered by many authors [2,10] but here we restrict  the  potential to a finite interval so as to treat a confined system and discuss the relationship of the solutions to the case where this one-dimensional potential is defined over the infinite interval.

\section{An example of a non-Hermitian, nonlinear operator}
Consider the following equation for a general radial potential $ V(r) $ defined on the space of square integrable  functions with the usual radial weight function $ r^2$
\begin{equation}
(-\frac{1}{2}D^2 -\frac{1}{r} D + V -s\ln(|\psi|))\psi = E \psi, s>0, D=\frac{d}{dr}
\end{equation}
The domain of the operator is the space of continuous and twice differentiable functions in this space and taking the inner product with $ \psi^{*} $ leads to real $ E$. Consequently this operator is appropriate for the bound states of the system.
The logarithmic potential ensures that this is a non-Hermitian and non-linear equation. If  we fix $ \psi $ , one of the solutions of (1) and consider the related equation:
\begin{equation}
(-\frac{1}{2}D^2 -\frac{1}{r} D + V -s\ln(|\psi|))\bar{\psi }= \bar{E} \bar{\psi} 
\end{equation}
then this is a separate equation where $ \bar{\psi} = \psi $ is a particular solution. Equation (2) is a linear , Hermitian system so that the eigenvectors are orthogonal and have a nodal structure. This structure is such that if we order the eigenvalues so that $ \bar{E_{n}}$  is the nth eigenvalue ( with $ \bar{E_{0}} $ being the minimum ) then the nodes interlace so that there is a node of the nth state between two successive nodes of the $ (n+1)st $ state. This general theorem applies only to Hermitian systems but it implies that $ \psi  $ has a finite number of nodes and there is an orthogonal  space   spanned by states of lower energy in (2)  each with a different number of nodes .This is true for all $s$  and as $ s \rightarrow 0 $ which corresponds to a Hermitian  system  the number of nodes in $ \psi$ does not change for all $ s$. Thus we can characterise the states by the number of nodes.
We now specialise to $ V(r)=-1/r $  in order to compare our calculations to an accurate set finite element calculations by Scott et al [13]. For any solution of (1)  there is a maximum node  so that for  $r$ greater than this maximum we may write 
\begin{equation}
\psi =exp(-\alpha(r) )
\end{equation}
which is nodeless.
 Substituting into (1) 
\begin{equation}
( (D \alpha)^2 -D^2(\alpha) +\frac{2}{r} D\alpha + \frac{2}{r} -2s\alpha +2E)exp(-\alpha)  = 0
\end{equation}
from which we may deduce
\begin{equation}
(1 -2s\alpha/ (D \alpha)^2-D^2(\alpha) /(D \alpha)^2 +\frac{2}{r} ( D\alpha )+1)/(D\alpha)^2  +2E/(D \alpha)^2 =0
\end{equation}
Equating  the first two terms  to zero we have
\begin{equation}
(D \alpha)^2 -2s\alpha=0  \Rightarrow \alpha = sr^2/2
\end{equation}
and, using (6), the  limit of the  remaining terms as $ r \rightarrow \infty $ is zero.Thus the leading asymptotic term is $ f= exp(-sr^2/2)$ so that 
\begin{equation}
\lim_{r\rightarrow \infty} \frac{\psi}{f} =1 \Rightarrow \psi \sim f  ,r \rightarrow \infty
\end{equation}
 We note  that asymptotic terms are not unique and  Scott [13]  uses an asymptotic form
 \begin{equation}
g = exp(-sr^2/2-br)= exp((-sr^2/2)(1+b/sr))
\end{equation}
 and we have the relationships
\begin{equation}
\psi/f \sim1, \psi/g \sim 1, f/g \sim 1 , r \rightarrow \infty
\end{equation}
When $r$ is sufficiently large so that $ \psi-f $ is negligible we can deduce properties of $ \psi $ in the asymptotic region. Essentially $ \psi>0, D\psi <0 , D^2\psi>0 $  Thus for all $r$ we can write
\begin{equation}
\psi=exp(-sr^2/2)\phi \Rightarrow ln|\psi | = -\frac{sr^2}{2} +ln |\phi | 
\end{equation}
and
\begin{equation}
(-\frac{1}{2} D^2-\frac{1}{r}(D+1) )\phi= (E-\frac{3s}{2} -rsD+s ln|\phi |)\phi
\end{equation}
which for $ s=0 $ reduces to the Schrodinger equation for atomic hydrogen.More generally, for the ground state we can solve this equation analytically by writing $ \phi= exp(-br) $ so that
\begin{equation}
-\frac{b^2}{2}+\frac{1}{r} (b-1) = (E-\frac{3s}{2} )
\end{equation}
Choosing $ b=1$ leads to 
\begin{equation}
\psi = exp(-sr^2/2-r), E=-\frac{1}{2}+\frac{3s}{2}
\end{equation}
Thus the ground state wave function is nodeless but to treat excited states, where the logarithmic term is more difficult to treat, we use iteration.We note that for all states if we renormalise $ \psi$ to $ N_{0}\psi$ and cancel $ N_{0} $ then the energy in (1) changes so that $ E \rightarrow E +sN_{0} $ Thus in order to compare our results with [12 ] we need a standard normalisation for the solutions. In the calculations we choose $ N_{0} $ so that
\begin{equation}
\int_{0}^{\infty} r^2|\psi|^2 dr =1
\end{equation}
Since the wave function is nodeless  this is the ground state solution.In order to calculate the excited states we need numerical methods in order to compare with the    extensive finite element methods in [13,14] . Here we use iteration by expanding the solution in terms of Laguerre functions. One of the major problems is the numerical integration of the logarithmic terms particular near the nodes of the wave function which will differ from the nodes of an approximate wave function.
We may reformulate (1) using the substitution $ x=\mu r =\sqrt{s}r $ which leads to 
\begin{equation}
(-\frac{1}{2} D_{x}^2 -\frac{1}{x} D_{x} -\frac {\mu}{x}-ln(|\psi|)\psi =\hat{E} \psi, \quad \hat{E}=E/\mu^2
\end{equation}

 The  iterative method, used is similar to Brillouin-Wigner  perturbation theory, together with a basis of generalised Laguerre polynomials  the form $ \{ L(m-1,1/2,cx^2) , m=1...\}$ which satisfy 
\[ \int_{0}^{\infty}L(m-1,1/2,cx^2)L(n-1,1/2,cx^2)x^2exp(-cx^2)dr =\]
\begin{equation}
< L(m-1,1/2,cx^2)L(n-1,1/2,cx^2) > = N_{n} \delta_{m,n}
\end{equation}
where $ N_{n} $ is an n-dependent normalisation constant and this defines an inner product for the $ \{ L(m-1,1/2,cx^2) \}$ which form a complete set in the Hilbert space defined by this inner product.Thus we approximate the nth  excited state by $ \psi_{n} =\phi _{n}exp(-cx^2/2) $ which is also in the space of square integrable functions defined earlier. The choice of $ c>0 $ is arbitrary but a choice may be made to improve the \emph{ rate of convergence} of the iterative procedure. We may write 
\begin{equation}
\psi_{n} = a_{n} (L(n-1,1/2,cx^2) + \sum_{m \neq n} a_{m} L(m-1,1/2,cx^2))exp(-cx^2/2)
\end{equation}
Substituting $ \psi_{n} $ into equation (15) leads to an equation for $ \phi_{n} $
\begin{equation}
H_{0} \phi_{n} - E_{n}\phi_{n}= - \hat{V}(\phi_{n}), \quad E_{n}=(\hat{E_{n}} +\frac{3c}{2})
\end{equation}
where
\begin{equation}
H_{0}= -\frac{1}{2}(D_{x}^2 +\frac{2}{x}D_{x}) +cxD_{x} \end{equation}
\begin{equation}
\hat{V}(\phi_{n}) = ( c(c-1)x^2/2-\frac{\mu}{x} -ln| \phi _{n}| )\phi_{n}
\end{equation}
The operator $ H_{0} $ is a Hermitian operator on the Hilbert space defined in (16) so that
\[< L(m-1,1/2,cx^2)H_{0}L(n-1,1/2,cx^2) >=\]
\begin{equation}
 < H_{0}L(m-1,1/2,cx^2)L(n-1,1/2,cx^2) >
\end{equation}
and 
\begin{equation}
H_{0}L(n-1,1/2,cx^2)=2c(n-1)L(n-1,1/2,cx^2)
\end{equation}
From equation (18) we construct an iteration scheme with $ a_{m} =0, m\neq n$ as initial function,  in the form
\begin{equation}
( H_{0}-E_{n}) \phi_{n}^{M+1} = (E^{1}\phi_{n}^{M}-V(\phi_{n}^{M})), E_{n} = 2c(n-1)
\end{equation}
Taking the inner product with $ L(n-1,1/2,cx^2) $ leads to
\begin{equation}
E^{1} = \frac{<L(n,1/2,cx^2)| V(\phi_{n}^{M})>}{a_{n}}
\end{equation}
which determines $ E^{1} $. Similarly taking the inner product with $ L(m-1,1/2,sx^2), m\neq n $ updates the wave function so that
\begin{equation}
2(m-n)ca_{m}^{M+1}  = <L(m,1/2,cx^2)| E^{1}\phi_{n}^{M}-V(\phi_{n}^{M})>
\end{equation}
Given the new coefficients we may renormalise, and hence define $ a[n]^{M+1} $ so that
\begin{equation}
\sum_{m=1..N} (a_{m}^{M+1})^2 =1
\end{equation}
Thus we have constructed an iteration process of the form
\begin{equation}
\mathbf{a} = F(\mathbf{a})
\end{equation}
In order to establish convergence or rapid convergence often an enhanced procedure such as Newton's method is needed.However we  used an effective, simpler procedure.For any $\nu, 0<\nu<1 $ we may rewrite(23) as
\begin{equation}
( H_{0}-E_{n}) \phi_{n}^{M+1} = \frac{1}{\nu} (E^{1}\phi_{n}^{M}-V(\phi_{n}^{M}))+\frac{1-\nu}{\nu}( H_{0}-E_{n}) \phi_{n}^{M+1}
\end{equation}
A new iteration procedure for the solution  may then be constructed  by replacing $ \phi_{n}^{M+1}$ on the right hand side of (28) by $\phi_{n}^{M}$ and choosing $ \nu  $ arbitrarily. An appropriate choice will be one that increases the rate of convergence. Convergence problems were also obtained in [13,14] and an additional problem is that the zeros of $\phi_{n}^{M} $ will differ and affect the numerical integration since $ \ln(|\phi_{n}^{M}|)$ is singular at these zeros. In the method presented here the logarithm is always multiplied by the function and $\phi_{n}^{M} \ln(|\phi_{n}^{M}|) $ is zero rather than infinite albeit that the derivative is infinite at these points.In order to estimate the integrals efficiently we use different  Gauss procedures, [15] ,between the zeros of $\phi_{n}^{M}$ at every stage. This technique is amplified in the appendix.In tables 1 and 2 we illustrate the eigenvalues and zeros of the eigenfunctions calculated compared with those from [13].For the first excited state we use 20 basis functions with $ c=1 $ for all except $ s=9,10 $ where we use $ c=0.5 $ and $ s=0.05 $ where $c=1.5 $. There are minor changes to the eigenvalues for the parameters in this range and the node is fairly stable. For the second excited state we use 25 basis functions with $ c=0.5 $ except for $ s=1,0.5$ where $ c=0.59 $ leads to enhanced convergence.The Finite Element Method (FEM) is subject to numerical integration errors and uses a maximum value of $ r$ . But the differences between the calculations reduces as the number of Laguerre basis functions increases.

\begin{table}[htp]
\caption{First Excited States}
\begin{center}
\begin{tabular}{|c|c|c|c|c|}
s&Eigenvalue&Zero&FEM-Eigenvalue&FEM-Zero \\
0.5&0.8485&1.4121&0.8463&1.4115\\
1&1.5002&1.1731&1.4982&1.1719\\
2&2.3874&0.9320&2.3859&0.9320\\
3&2.9416&0.8034&2.9396&0.8034\\
4&3.2719&0.7181&3.2687&0.7183\\
5&3.4325&0.6562&3.4278&0.6566\\
6&3.4561&0.6085&3.4493&0.6091\\
7&3.3648&0.5703&3.3558&0.5710\\
8&3.1745&0.5387&3.1668&0.5395\\
9&2.8867&0.5133&2.8829&0.5129\\
10&2.5289&0.4903&2.5251&0.4900
\end{tabular}
\end{center}
\label{default}
\end{table}%

\begin{table}[htp]
\caption{Second Excited States}
\begin{center}
\begin{tabular}{|c|c|c|c|c|c|c|}
s&Eigenvalue&Zero-1&Zero-2&FEM-Eigenvalue&FEM-Zero-1&FEM-Zero-2\\
0.5&1.3291&1.2903&3.8422&1.3258&1.2818&3.819\\
1&2.3437&1.0565&3.0123&2.3390&1.0537&3.0018\\
2&3.9231&0.8336&2.2936&3.9190&0.8326&2.2896\\
3&5.1520&0.7154&1.9331&5.1478&0.7146&1.9334\\
4&6.1480&0.6382&1.7062&6.1429&0.6377&1.7083\\
5&6.9685&0.5825&1.5459&6.9622&0.5822&1.5490\\
6&7.6479&0.5397&1.4241&7.6401&0.5396&1.4285\\
7&8.2093&0.5054&1.3287&8.1997&0.5055&1.3328\\
8&8.6626&0.4777&1.2534&8.6576&0.4775&1.2534\\
9&9.0321&0.4539&1.1876&9.0261&0.4536&1.1890\\
10&9.3250&0.4332&1.1303&9.3158&0.4332&1.1330

\end{tabular}
\end{center}
\label{default}
\end{table}%

\section{Examples of non-Hermitian operators with complex terms}

Here we consider the following operator defined on the Hilbert space of least squares space in $ -T <x<T $ where the domain is the subset of continuous  and twice differentiable functions in this space and 
\begin{equation}
H \psi =(-D^2 + iV(x) )\psi = E \psi  , \psi(\pm T ) =0, \quad D= \frac{d}{dx}
\end{equation}
and where $ V(x) $ is an \emph{odd} function of x. The domain can be more precisely defined by the basis functions
\begin{equation}
\{ cos(\frac{(2n-1)\pi x}{2T}) , sin(\frac{n\pi x}{T} )\}, n=1,2...
\end{equation}
which form a complete set for the subspace of  the Hilbert space  where the elements of the space vanish at $ x=\pm T $.  This is the set of functions in the Hilbert space orthogonal to $ 1 $  and it is also a formalism for  $ V(x) $ a continuous and differentiable   piecewise function  The solution may be represented in this basis and this may also be a representation converging in the mean  to a piecewise  continuous and differentiable function so that, at points of  finite discontinuity of the solution or the derivative, the average of the two values is obtained.In all cases we have  finite values of the second derivative.

Of course we can find different eigenvalues with a different set of boundary conditions such as the derivatives vanishing at $ x= \pm T $ or a a more generic domain that encompasses both types of solution.So that the eigenvalues depend on the definition of the  chosen operator, which includes the domain and range.

Equation (29) satisfies PT symmetry which implies that either $ E$ is real (unbroken PT ) or we have double degenerate states where $ E $ is complex (broken PT symmetry).
If we now assume that the solution can be written in the form
\begin{equation}
\psi = \phi_{1} + i \phi_{2} , \quad \phi_{1},\phi_{2}  \quad   real
\end{equation}
with  $ \phi_{1} $ even and $ \phi_{2} $  odd then 
\[ < \psi| H\psi> = E <\psi|\psi> \Rightarrow\]
\begin{equation}
\int_{-T}^{T}( -(\phi_{1}D^2\phi_{1} + \phi_{2}D^2\phi_{2} ) )dx = E\int_{-T}^{T} (\phi_{1}^2 + \phi_{2}^2)
\end{equation}
so that $E $ is real and we have unbroken PT. Small perturbations of $ \phi_{1} , \phi_{2} $ so that the odd and even symmetries are broken lead to complex values for E. So that we can deduce that (32) hold for real $ E $ unless there is some \emph{accidental } choice of parameters in $ V(x) $ that lead to additional real eigenvalues.

As an example we treat $ V(x) = x $.Normalising the basis functions and using variational theory with $N $ of these functions we solve the finite matrix eigenvalue problem where the non-zero elements of the matrix $ h$ are 
\begin{equation}
h_{n,n} = \frac{n^2\pi^2}{4T^2}, h_{2n-1,2n}=h_{2n,2n-1} = \frac{i}{T} \int_{-T}^{T} x\omega_{2n-1}\omega_{2n}dx
\end{equation}
where 
\begin{equation}
\omega_{2n-1} = cos(\frac{(2n-1)\pi}{2T} , \omega_{2n}=sin(\frac{2n\pi x}{2T} )
\end{equation}
Since the basis functions are bounded the off diagonal elements are $ O(T) $.Thus for fixed $ T $ and large $n$ the corresponding row of the matrix is diagonally dominant.The behaviour of the eigenvalues of $ h$ as $ T $ and $ N $  varies can be illustrated by some simple calculations. In table we fix $ N=4 $ for increasing $ T$
\begin{table}[htp]
\caption{Real parts of the eigenvalues of h ($N=4 $)}
\begin{center}
\begin{tabular}{|c|c|c|}
T=1&T=3& T=4\\
2.485&1.168&1.105\\
9.864&1.1.68&1.105\\
22.203&2.433&1.208\\
39.469&3.446&1.208
\end{tabular}
\end{center}
\label{default}
\end{table}%
The repeated elements in the columns indicate the complex pairs and as $ T $ increases two critical points are reached and for $ T=4$ all the eigenvalues are complex pairs. However if we fix $ T $ and increase $ N $ the diagonal dominance causes an additional behaviour as illustrated in table .
\begin{table}[htp]
\caption{Real parts of the eigenvalues of h ($ T=5 $)}
\begin{center}
\begin{tabular}{|c|c|c|c|}
N=4& N=8&N=16&N=32\\
0.7309&1.1697&1.691&1.691\\
0.7309&1.1697&1.691&1.691\\
0.7495&2.0292&2.0439&2.0439\\
0.7495&2.0292&2.0439&2.0439\\
-&2.5577&2.8539&2.8539\\
-&2.5577&2.8539&2.8539\\
-&4.3104&4.4418&4.4418\\
-&4.3104&4.4418&4.4418\\
-&-&6.0061&6.0061\\
-&-&7.7448&7.7448

\end{tabular}
\end{center}
\label{default}
\end{table}%
As illustrated in table for increasing  $ N $ we obtain convergence for the lower states, but for $ n>8 $ all  the eigenvalues are real. This  arises when $ N >> T $ and final rows in the corresponding matrix exhibit diagonal dominance with the diagonal elements increasing relative to the matrix elements. The form of the matrix when $ V=x^3 $ is similar excerpt that the non-diagonal elements are $ O(T^3) $. For $ T=15, N=50 $  the  lowest  10 real parts of the eigenvalues are:
\[ 1.156,4.109,7.562, 11.314,15,294,19.450,23.773,28.088,32.858,32.858 \]
These are close to the values given in  [ 2]  for the infinite interval.  For these parameters most of the values listed  are real to the accuracy of the calculation , but the last two eigenvalues  form a complex pair. However for larger N these split to form two real eigenvalues.

\subsection{The relationship between the finite and infinite interval}

In order to examine how these  problems are related to similar problems for an infinite interval ( when $ T \rightarrow \infty $) we consider $ V=x^m$ ( $m$ odd)   in the Hilbert space of square integrable  functions in $ -\infty <x < \infty  $ so that $H=-D^2 +ix^m $ where the domain is the space of continuous and differentiable functions, (including  piecewise continuous and differentiable continuous functions) so that the second derivative is finite  . For a more precise definition of the operator we need to define the asymptotic boundary conditions analogous to the wave function being zero at $ x=\pm T $ in the finite interval. Essentially the asymptotic analysis replaces the confinement in a finite interval.There are many possibilities for the behaviour at $ \pm \infty$ but we require the wave function to be normalisable. One possibility is where  for  $ x>0 $, the wave function can be expressed in the form 
\begin{equation}
\psi =exp(-bx ^p)\psi_{1},   Re(b)>0 
\end{equation}
where $ \psi $ is continuous and differentiable throughout the region which implies that $ D^{2}\psi $ is finite  
and where  $ \psi_{1} $ is $ O(x^q) $  for some real  $ q $ as $ x\rightarrow \infty $ Thus we have 
\begin{equation}
(-D^2+ ix^m)\psi =E \psi,  -\infty < x< \infty  \end{equation}
which implies that
\[ (-D^2\psi_{1} + bp(p-1) x^{p-2}\psi_{1} +2bpx^{p-1}D\psi_{1} -b^2p^2x^{2p-2}) \psi_{1} ) + \]
 \begin{equation} ix^m \psi_{1} - E\psi_{1} = 0
\end{equation}
Choosing
\begin{equation}
2p-2 = m , b^2p^2 =i
\end{equation}
eliminates the potential term $ ix^m $ and leads to $ p=(m+2)/2 $ and $ b= \frac{2}{m+2} \sqrt(i).$ This implies
\begin{equation}
b= \frac{2}{m+2} ( cos(\pi/4) + isin(\pi/4 ) = \frac{\sqrt{2}}{m+2} (1+i)
\end{equation}
since $ Re(b) >0 $.The equation for $ \psi_{1} $ becomes
\begin{equation}
(-D^2 + b\frac{m^2+2m}{4} x^{(m-2)/2} +(m+2)bx^{m/2}D -E)\psi_{1} =0
\end{equation}
We now express $ \psi_{1} $ in the form 
\begin{equation}
\psi_{1} = x^q ( \sum_{n=0} ^{\infty} \frac{A_{n}}{x^n} ), \quad A_{0} =1
\end{equation}
for an analysis of the asymptotic behaviour. Dividing by $ x^{q+(m-2)/2} $  we have the leading term 
\begin{equation}
\frac{m^2+2m}{4}b +(m+2)bq
\end{equation}
For this to be zero we require  $ q= -m/4$  so that finally we have as $  x>0, x\rightarrow  \infty$
\begin{equation}
\psi \sim exp(-\frac{\sqrt{2}}{m+2} (1+i)|x|^p)|x|^{-m/4} 
\end{equation}
where for $ x>0 $, x is identical with $ | x | $. 
Thus $ \psi \rightarrow 0 $ as $ x \rightarrow \infty$.  In the equivalent  calculation in $ x<0 $ by putting  $ y=-x $ we have the same equation in y except that $ i \rightarrow -i $  so that the analysis is for $ \sqrt{-i}$. An asymptotic solution for $ x<0 $ is therefore
\begin{equation}
\psi \sim exp(-\frac{\sqrt{2}}{m+2} (1-i)|x|^p)|x|^{-m/4} 
\end{equation}
We note that  solutions may be found with these asymptotic conditions by replacing the variable $ x $ by the complex variable $ z $ and using the analysis of Stoke lines to match the solutions. However here we restrict the analysis to the real variable $ -\infty< x< \infty $ and the operators are defined on this space.

The asymptotic  conditions are consistent with $ \psi $ being square integrable but we also require that the second derivative to be finite in the domain  as $ x \rightarrow 0 $. To ensure this we use a more appropriate  basis for the infinite interval in the form:
\begin{equation}
\phi_{n} = exp(-\alpha x^2/2) \hat{\omega}_{n} (\sqrt{\alpha} x)N_{n}, N_{n}=\frac{1}{2^2n!}\quad \sqrt{\alpha /n}
\end{equation}
where $ \hat{\omega}_{n} $ are the nth degree Hermite polynomials and , for any $ \alpha >0 $ is a dense, complete set in this space. For a finite set of these functions , $ n=1...N $ we may represent the system as a finite matrix using
\begin{equation}
x\phi_{n} = (\sqrt{n+1}\phi_{n+1} + \sqrt{n} \phi_{n-1} )/\sqrt{2\alpha}
\end{equation}
and
\begin{equation}
-\frac{d^2\phi_{n}}{dx^2} =( -\alpha x^2 + \alpha (2n+1))\phi_{n}
\end{equation}
Considering the case $ V=x^3 $  we note that there  is an arbitrary scale $ x \rightarrow \gamma x $ so that we need to solve a matrix eigenvalue problem of the form
\begin{equation}
\hat{h} =(h_{N} +i\gamma^5 v)\mathbf{x} = E_{0} \mathbf{x} , \quad  E_{0} = E\gamma^2
\end{equation}
Taking $ \alpha=1, \gamma = 0.5 $ and $ N=90 $ leads to the following results for the first $ 10 $ eigenvalues.\begin{table}[htp]
\caption{Lower eigenvalues for $V=x^{3}$}
\begin{center}
\begin{tabular}{|c|c|}
1.156267&$1.04.10^{-21}$\\
4.109229&$9.76.!0^{-21}$\\
7.562274&$1.87.10^{-19}$\\
11.314422&$8.27.10^{-19}$\\
15.291554&$3.56.10^{-18}$\\
19.451529& $3.36.10^{-17}$\\
23.766740&$5.76.10^{-17}$\\
28.217525&$6.1.10^{-16}$\\
32.789083&$1.18.10^{-15}$\\
37.469825&$7.36.10^{-15}$
\end{tabular}
\end{center}
\label{default}
\end{table}%
These values agree with those given in [2,10] up to the digits quoted. But in addition there are complex eigenvalues. For example $ 20.37329 \pm i247.169708 $. It is well known that the eigenvalues of this system are real. But for any $ N $ there is a  value of  $ T $  so that the absolute value of the approximate wave function in this basis  is essentially zero in $ |x|  > T  $. Thus  the calculation is essentially  a confined calculation. Increasing $ N $ changes $ T$ and  the complex states significantly and we can derive a process for estimating if the value of $ N $ leads to the real eigenvalues.Suppose we solve the matrix eigenvalue problem for $ \hat{h} $ for a particular state obtaining an approximate eigenvalue $ E_{0}  $ and use the normalised eigenvector to form an approximation in the form
\begin{equation}
\hat{\psi} = \sum_{n=1}^{N} a_{n} \psi_{n}
\end{equation}
where \begin{equation}
H\hat{\psi} =\sum_{n=1}^{N} \sum_{m=1}^{N}a_{n}H_{mn} \phi_{m} + \sum_{n=1}^{N} \sum_{m=N+1}^{M+N}H_{mn} a_{n}\phi_{m} 
\end{equation}
where $ M=3 $ since the matrix is band-limited. We note that the two terms in this equation are orthogonal. We denote the second term by $ \chi$  and if $ \chi=0 $ then $ \hat{\psi} $ is exact.  Thus for the exact eigenvalue $ E $, since the matrix elements  $ \hat{h}_{mn} =H_{mn} , 1\leq n,m \leq N $  we have 
\begin{equation}
<(H-E) \hat{\psi} | (H-E) \hat{\psi}>= (E_{0}-E)^2 + <\chi | \chi> =(E_{0} -E)^2 + \Delta
\end{equation}
Thus $ \Delta $ can be calculated and if the value is small we can accept the approximation whereas a large value will imply that the size of the basis is not sufficient to estimate this state. The values of $ \Delta $ for the states are given in the table indicating that the values are good approximations whereas for $ E_{0}=20.37329 \pm i247.169708 $ we have $ \Delta = 87.829 $ so that the basis is not sufficient for the exact value of that state.

\section*{Appendix: Evaluation of the integrals}
For the first problem, the non-Hermitian , non-linear potential, numerical integration is required. At every stage we choose formulae between the zeros of $\phi_{n}^{M} \ln{\phi_{n}^{M}} $ so that the integrands are not singular. The derivatives of the integrands are singular at these zeros but the standard error term is defined  between these points. Although this can theoretically lead to  large error values  the results were consistent. The method used was a version of Gauss integration with a given weight function $ w(x) $  with the construction of orthogonal, monic functions  $ p_{k} , k=0..n_{1}  $ [15], so that 
\begin{equation}
\int_{a}^{b} p_{k}(x)p_{m}(x) w(x) dx = 0, m\neq k
\end{equation}
We use both the finite and infinite interval ($ b=\infty$ ) in our calculations and the general formula for the  the integrals used is 
\begin{equation}
\int_{a}^{b} f(x)w(x) dx =\sum_{j=1}^{n_{1}} W_{j}f(x_{j} )
\end{equation}
where the $ x_{j} $ are the zeros of $ p_{n_{1}}(x) $ and the weights are given by
\begin{equation}
W_{j} = \frac{\int_{a}^{b}w(x)p_{n1-1}(x)^2 dx}{p_{n_{1}-1}(x_{j})Dp_{n1}(x_{j})}
\end{equation}
Here we use $ w(x) = exp(-cx^2) ,p_{0} =1,$ and choose d so that $p_{1}=x-d$ is orthogonal to $ p_{0} $. Subsequently we use the recurrence relation
\begin{equation}
p_{k+1} = (x+\alpha_{k})p_{k} + \beta_{k} p_{k-1} , \gamma_{k} =\int_{a}^{b} p_{k}^2w(x) dx
\end{equation}
where using the orthogonality of the constructed functions $ p_{n} , n<k+1 $,
\begin{equation}
\alpha_{k} = - \int_{a}^{b} xp_{k}^{2} wdx/\gamma_{k} , \beta_{k} =- \int_{a}^{b} xp_{k}p_{k-1} wdx/\gamma_{k-1} =\gamma_{k}/\gamma_{k-1}
\end{equation}
The simplification of $ \beta_{k} $ follows from the orthogonality condition and the the choice that the polynomials are monic. Simplifications can also be obtained for $ \alpha_{k}$ and $ \gamma_{k} $ by integration be parts.
\begin{equation}
\gamma_{k} =(k/2c)\gamma_{k-1} -\frac{1}{2c} [exp(-cx^2)p_{k-1}p_{k}]^{b}_{a}
\end{equation}
and
\begin{equation}
\alpha_{k}= \frac{1}{2c\gamma_{k}} [exp(-cx^2)p_{k}^2]^{b}_{a}
\end{equation}

The integrals required for the confined problem are analytic but they can be evaluated quickly by substituting the required angles in the result
\begin{equation}
Im\{\int x^m( exp(i(A+B)x ) + exp(i(A-B)x)dx \}= 2\int x^m(cos(Bx)sin(Ax) dx
\end{equation}
\section{References}
[1] Moiseyev N,2011, "Non-Hermitian Quantum Mechanics", Cambridge University Press, Cambridge. \newline
[2] Bender C M,2007, "Making sense of non-Hermitian Hamiltonians". Reports on Progress in Physics. 70 (6), 947; arXiv: hep-th/0703096  \newline
[3] Bender C M and  Boettcher S, 1998, "Real Spectra in Non-Hermitian Hamiltonians Having $PT$ Symmetry". Physical Review Letters. 80 (24);arXiv:physics/9712001 \newline
[4] Mostafazadeh A, 2002, "Pseudo-Hermiticity versus symmetry: The necessary condition for the reality of the spectrum of a non-Hermitian Hamiltonian". Journal of Mathematical Physics. 43 (1), 205; arXiv:math-ph/0107001 \newline
[5] Mostafazadeh A, 2002, "Pseudo-Hermiticity versus PT-symmetry. II. A complete characterization of non-Hermitian Hamiltonians with a real spectrum". Journal of Mathematical Physics. 43 (5): 2814; arXiv:math-ph/0110016 \newline
[6] Znojil M, 2001, Phys.Lett.A285, 7 \newline
[7] Znojil M, 2005, J.Math.Phys.46 062109 \newline
[8] Znojil M and Levai G, 2001, Mod.Phys.Lett.A16, 2273 \newline
[9]   Bender C,  Brody D C, Jones, H 2002, Phys. Rev. Lett. 89,270401  \newline
[10] Burrows B L and Cohen M ,2021, Eur.Phys J. D  75, 70 \newline
[11] Feinberg J, 2011,International Journal of Theoretical Physics volume 50, 1116 \newline
[12] Bagarello F and Fring A, 2013, Phys.Rev. 8, 042119 ; arXiv.1310.4775v[quant-ph]  \newline
[13]  Scott T C  and  Shertzer J ,2018, J. Phys. Commun. 2 , 075014 \newline
[14]  Shertzer J and  Scott T C, 2020, J. Phys. Commun. 4, 065004 \newline
[15] Steen N M ,Bryne G D and Gelbard E,1969, Mathematics of Computation, 23, 661

\end{document}